\documentstyle[psfig,aps,amssymb]{revtex}

\begin{document}
\draft

\title{
 	  A variational principle for time of arrival of null geodesics
       }

\author{Simonetta Frittelli$^{a,b}$\thanks{e-mail: 
	simo@mayu.physics.duq.edu}
	\and
	Ezra T. Newman$^b$
}

\address{
	$^a$ Physics Department, Duquesne University, Pittsburgh, PA 15282\\
	$^b$ Department of Physics and Astronomy, University of Pittsburgh
	 Pittsburgh, PA 15260.		
	}

\date{\today}

\maketitle

\begin{abstract}
Normally the issue or question of the time of arrival of light rays at an
observer coming from a given source is associated with Fermat's Principle of
Least Time which yields paths of extremal time. We here investigate a
related but different problem. We consider an observer receiving light from
an extended source that has propagated in an arbitrary gravitational field.
It is assumed from the start that the propagation is along null
geodesics. Each point of the extended source is sending out a light-cones
worth of null rays and the question arises which null rays from the source
arrive first at the observer. Stated in an a different fashion, a pulse of
light comes from the source with a wave-front as the leading edge, which
rays are associated with that leading edge. In vacuum flat-space we have,
from Huygen's principle, that the rays normal to the source constitute the
leading edge and hence arrive first at an observer. We here investigate this
issue in the presence of a gravitational field. Though it is not obvious,
since the rays bend and are focused by the gravitational field and could
even cross, in fact it is the normal rays that arrive earliest. We give two
proofs both involving the extemization of the time of arrival, one based on
an idea of Schrodinger for the derivation of gravitational frequency shifts
and the other based on V.I. Arnold's theory of generating families.

\em Dedicated to Jayant Narlikar. 
\end{abstract}


\section{Introduction \label{sec:1}}


If one considers the problem of the ``time of arrival'' of a light signal
coming from an extended source (a two-surface) and arriving at an observer
who is moving along an arbitary time-like world-line, one normally invokes
Fermat's Principle of Least Time to determine the path of the light-ray from
each point on the source to the observer. We will discuss a different but
related issue - at each source point there is a light-cones worth of rays
that are emitted and the question is what are the directions of the rays for
them to have the earliest ``time of arrival'' at the observer. Stated in a
different manner, if the source suddenly lights-up, then from each point
there will be a sphere's worth of rays leaving it resulting in a pulse
arriving at the observer from different source points with a leading edge
followed by a tail. The question then is which rays arrive earliest? From
the start we assume that the paths of the rays are given by null geodesics,
so that there is no need to invoke Fermat's Principle. In flat vacuum
space-time it is clear that it is the normal rays that arrive earliest,
i.e., they travel the shortest distance. However in a Lorentzian manifold
with a time varying metric the issue is not as obvious. The problems are
that there is no well defined distance between source and observer, the rays
can bend and be focused by the field and could even cross or even more than
one ray could arrive at the observer from the same source point.
Nevertheless we will show that it is the rays that are emitted normal to the
source that are the ones that arrive earliest, i.e., the time of arrival
function is extremized by the normal rays.

We will give two alternative proofs of this. The first proof, given in
Sec.II, involves a modification of the beautiful derivation by Schrodinger
of the gravitational frequency shift while the second proof, in Sec.III,
uses the theory of generating families of V.I. Arnold.

Probably the most powerful technique for the study of wavefronts and their
related characteristic (or null) surfaces in Lorentzian space-times and the
associated difficulties in the analytic description of the development of
caustics and crossover regions is Arnold's theory of Lagrangian and Legendre
submanifolds and the associated Lagrange and Legendre maps\cite{yellowarnold}%
. One of the main ingredients in this theory is the construction of what has
been referred to as generating families. They are, in general, two-point
functions, F(x,s) (chosen from, perhaps, different spaces, x and s, with
perhaps different dimensions), that are to be constructed from physical
arguments and which are stationary with respect to variations in one of the
two different spaces. In our case we give an example of this construction
where the two-point function is the time-of-arrival function of light rays
which begin from points on a two-surface, embedded in a four dimensional
space-time, (thought of as a source of radiation), and which end at points
on a one-dimensional manifold, a curve (thought of as the world-line of an
observer of that radiation) also embedded in the same four-space. Though we
will not amplify on it here, this example appears to be, in principle if not
in practice, of generic use in the theory of gravitational lensing in any
Lorentzian space-time.


\section{The Time of Arrival Function}

\label{sec:2}


We are concerned with the travel time of light signals from an extended
source to a localized observer. For our purposes, the source lights up
instantaneously, in its own rest frame, emitting photons in all directions
from every point on its (closed) surface. There is one photon that arrives
first at the observer's location, in the observer's proper time. If the
metric is stationary, then this photon is, intuitively, the one that takes
the shortest spatial path, perpendicularly to the surface of the source. In
the following, we make these notions more precise, extending them to the
case of arbitrary metrics.

Consider, in an arbitrary Lorentzian four-dimensional manifold, a given
closed space-like two-surface, ${\cal {S}}$, described by 
\begin{equation}
x^{a}=x_{0}^{a}(s^{1},s^{2})
\end{equation}
where $x^{a}$ are space-time coordinates in the neighborhood of the source,
and $s^{J}=(s^{1},s^{2})$ parametrize the surface ${\cal {S}}$. In addition,
consider a timelike worldline, ${\cal {L}}$. In the neighborhood of the
worldline, with no loss of generality, let the local coordinates be such
that ${\cal {L}}$ is given by $(\tau ,X^{i})$ where the $X^{i}$ are three
constants, the spatial location of the observer, and $\tau $ is the proper
time along the worldline.

From each point $s^{J}$ of ${\cal {S}}$, construct its future
lightcone, ${\cal C}_{s^{J}}$. In general, in the absence of horizons, the
line ${\cal {L}}$ intersects each ${\cal C}_{s^{J}}$ at least once. The
intersection takes place at a particular value of the proper time $\tau $
for each point $s^{j}$ on the surface. This means that there is a two-point
function 
\begin{equation}
\tau =T(X^{i},s^{J}).  \label{propertime}
\end{equation}
that represents the proper time of arrival at ${\cal L}$ of light signals
from ${\cal S}$. One explicit way of constructing such a function is as
follows. The lightcone ${\cal C}_{s^{J}}$ is foliated by light-rays from the
point $x_{0}^{a}(s^{J})$, which are solutions 
\begin{equation}
x^{a}=\gamma ^{a}(r;s^{J},\theta ,\phi )
\end{equation}
of the geodesic equation with initial data labeled by the initial point $%
s^{J}$ and the initial direction $(\theta ,\phi )$ of the ray. Here $r$ can
be thought of as an affine parameter along the null geodesics. The
intersection of the lightcone with the worldline ${\cal L}$ takes place at
points where 
\begin{equation}
\gamma ^{i}(r;s^{J},\theta ,\phi )=X^{i}  \label{inters}
\end{equation}
and the time ($x^{0}=\tau $) at which the lightray reaches the observer is 
\begin{equation}
\tau =\gamma ^{0}(r;s^{J},\theta ,\phi )  \label{time}
\end{equation}
where the values of $(r;s^{J},\theta ,\phi )$ in the right-hand side are
restricted by (\ref{inters}). In cases where (\ref{inters}) is invertible
for every value of $s^{J}$, it provides $(r,\theta ,\phi )$ as functions of $%
(X^{i},s^{J})$, which can be inserted into (\ref{time}) to yield (\ref
{propertime}).

\begin{figure}[tbp]
\centerline{\ 
\hbox{\vbox{\psfig{figure=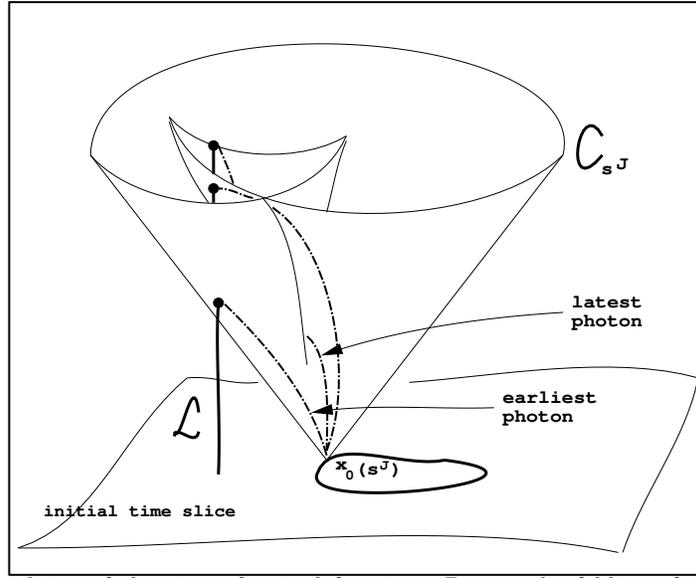,height=3in,angle=0}}
	} }
\caption{Origin of the multiplicity of the time of arrival function. Due to
the folding of the lightcone in the presence of curvature, there are several
null geodesics that reach the same spatial location, at different times. In
our picture, three null geodesics, leaving $x_0$ in different directions at
the same time, reach the same worldline.}
\label{fig:multiplicity}
\end{figure}

For large distances between the source and observer, a worldline intersects
any generic future lightcone several times, due to the folds in the
individual light-cones produced by space-time curvature (see Fig.~\ref
{fig:multiplicity}). This is the case where (\ref{inters}) is not
invertible, since for every fixed value of $s^{J}$ there would be several
values of the set $(r,\theta ,\phi )$ corresponding to the same spatial
location $X^{i}$. This means that there are several photons, shot in
different directions, that reach the observer's location at different times.
Therefore, for large distances the function $T(X^{i},s^{J})$ is multivalued.
We restrict attention to such cases where $T(X^{i},s^{J})$ is single valued.
In this case, there exists the following theorem, mentioned in the special
case of static space-times by Arnol'd (see~\cite{yellowarnold}, p. 251, and~%
\cite{greenarnold}, p. 298):

\noindent Theorem. {\em The proper time of arrival, $T$, at ${\cal L}$, is
extremized by those rays that leave ${\cal S}$ perpendicularly to it. In
other words, the points $s^{J}$ such that 
\begin{equation}
\frac{\partial T}{\partial s^{J}}(X^{i},s^{J})=0
\end{equation}
are connected to ${\cal L}$ by light-rays that are normal to ${\cal S}$. }

\noindent{\bf Proof:} The proof is based on the standard variational
principle for null geodesics, and is an extension of a similar result in~%
\cite{schrodinger}. Consider the action 
\begin{equation}
I(p,q) = \int_0^1 g_{ab}\dot{x}^a\dot{x}^b dr,
\end{equation}
where $\dot{x}^a\equiv dx^a/dr$ is the tangent vector to an affinely
parametrized null geodesic between the points $p=x^a(0)$ and $q=x^a(1)$, and 
$s$ is the affine parameter. Consider the variation of $I$ constructed by
taking the difference between two neighboring null geodesics with different
initial points, $p_1$ and $p_2$, and different final points, $q_1$ and $q_2$
Since $I$ evaluates identically to zero in both instances, its variation is
zero as well, 
\begin{equation}  \label{zero}
0= \Delta I = I(p_2,q_2) - I(p_1,q_1)
\end{equation}
The variation is 
\begin{eqnarray}
\Delta I &= & \int_0^1 g_{ab,c}\dot{x}^a\dot{x}^b\delta x^c + 2 g_{ab}\dot{x}%
^a\delta \dot{x}^b \; dr  \nonumber \\
&= & \int_0^1\left( (g_{ab,c}-2 g_{cb,a})\dot{x}^a \dot{x}^b -2g_{cb}\ddot{x}%
^b \right)\delta x^c + 2 \frac{d}{dr} \left(g_{ab}\dot{x}^a\delta
x^b\right)\;dr . \\
\end{eqnarray}
Since the curves are null geodesics, the term proportional to $\delta x^c$
in the integrand vanishes, and we are left with 
\begin{equation}  \label{variation}
\Delta I = 2 \left.\left(g_{ab}\dot{x}^a\delta x^b\right)\right|^1_0.
\end{equation}
By (\ref{zero}) and (\ref{variation}), we have 
\begin{equation}  \label{trans}
\left. g_{ab}\dot{x}^a \delta x^b \right|_{r=1} = \left. g_{ab}\dot{x}^a
\delta x^b \right|_{r=0}
\end{equation}
In (\ref{trans}), $\delta x^b$ represents an arbitrary (up to the condition
that $p$ and $q$ can be connected by a null geodesic) displacement at $r=0$
and $r=1$ between the null geodesic with tangent $\dot{x}^b$ and a
neighboring one. We now particularize (\ref{trans}) to our case of interest,
in which, initially, neighboring null geodesics are connected by
displacements on the surface ${\cal S}$; {\em i.e.\/}; 
\begin{equation}  \label{initial}
\delta x^b|_{r=0} = \left( \frac{\partial x^b_0}{\partial s^J}\right) ds^J
\end{equation}
where $\frac{\partial x^b_0}{\partial s^J}$ are the two coordinate tangent
vectors to ${\cal S}$ and $ds^J$ is arbitrary. The final displacement must
be tangent to ${\cal L}$, {\em i.e.\/}, 
\begin{equation}  \label{final}
\delta x^b|_{r=1}= v^b d\tau
\end{equation}
with $v^b$ the tangent vector to the curve ${\cal L}$. However, because the
two null geodesics arriving at $r=1$ and separated by $\delta x^b|_{s=1}$
must be the same pair of null geodesics leaving $r=0$ separated by $\delta
x^b|_{r=0} $ then $d\tau$ is not arbitrary, but 
\begin{equation}  \label{finaltau}
d\tau = dT|_{X^i} = \frac{\partial T}{\partial s^J} ds^J.
\end{equation}
With (\ref{initial}), (\ref{final}) and (\ref{finaltau}), Eq.~(\ref{trans})
reads 
\begin{equation}  \label{almost}
\left( g_{ab}\dot{x}^a \frac{\partial x^b_0}{\partial s^J} - g_{ab}\dot{x}^a
v^b \frac{\partial T}{\partial s^J} \right) ds^J = 0.
\end{equation}
Since $ds^J$ is arbitrary, and since $g_{ab}\dot{x}^a v^b$ can not vanish as
long as $\dot{x}^a$ and $v^b$ are tangent to a null and a timelike curve,
respectively, then (\ref{almost}) is equivalent to 
\begin{equation}
\frac{\partial T}{\partial s^J} = \left(\frac{1} {g_{cd}\dot{x}^c v^d}%
\right) g_{ab}\dot{x}^a \frac{\partial x^b_0}{\partial s^J} .
\end{equation}

This implies that, at each point of ${\cal {S}}$, $\partial T/\partial s^J$
vanishes if and only if the null ray $\dot{x}^a$ is normal to the surface at
that point. This proves the theorem.$\Box$

Note that since no property of the line ${\cal L}$ was used, the theorem can
be restated as follows. Given a time foliation of a Lorentzian manifold with
local coordinates chosen as $(\tau ,X^{i})$, and a source, a closed
two-surface that ``lights up'', the time $\tau =T(x^{i},s^{J})$ of arrival
at any spatial point $X^{i}$, of light signals from a surface point $s^{J}$,
is extremized by the light-rays leaving the surface perpendicularly.


\section{Eikonals and the Time of Arrival}

\label{sec:3}


In the following, we provide an alternative method for obtaining the time of
arrival function which is based entirely on the use of the eikonal equation
- with Arnold's generating families - and specifically on knowledge of a
two-parameter family of solutions of the eikonal equation,

\begin{equation}
g^{ab}(x^{c})\partial _{a}Z\partial _{b}Z=0;  \label{a}
\end{equation}
i.e., it is assumed that a solution, with the two parameters $\alpha
^{A}=(\alpha ^{1},\alpha ^{2})$

\begin{equation}
u=Z(x^{a},\alpha ^{A})  \label{b}
\end{equation}
to Eq.(\ref{a}) is known. Then for each value of $\alpha ^{A}$ the level
surfaces of $Z$ are null (i.e., $\partial _{a}Z$ is a null covector).
Furthermore it is assumed that at each point $x^{c},\partial _{a}Z$ sweeps
out the entire null cone at $x^{a}$ as $\alpha ^{A}$ goes through its range.

\vspace{0.4cm} \noindent{\bf Remark.} We point out and emphasize that the
level surfaces of the solutions to Eq.(\ref{a}) though referred to as ``null
or characteristic surfaces'' are not strictly speaking surfaces; they can
have self-intersections and in general are only piece-wise smooth. Though
Arnold refers to them as ``big-wave-fronts'' we will continue to call them
null surfaces. The intersection of a big wave front with a generic three
surface yields a two-dimensional (small) wave front.

The first thing that we want to show is that the light-cone, ${\frak C}
_{x_{0}},$ from an arbitrary space-time point $x_{0}^{a}$ can be constructed
from knowledge of the function $Z$ of Eq.(\ref{b}).\qquad

One sees immediately, from Eqs.(\ref{a}) and (\ref{b}), that the function 
\begin{equation}
S^{*}(x^{a},x_{0}^{a},\alpha ^{A})=Z(x^{a},\alpha ^{A})-Z(x_{0}^{a},\alpha
^{A})=0  \label{c}
\end{equation}
defines a two-parameter set of surfaces which all pass thru the point $%
x_{0}^{a}$ and which, furthermore, are all null surfaces. The envelope of
this family is constructed by demanding that 
\begin{equation}
\partial _{A}S^{*}(x^{a},x_{0}^{a},\alpha ^{A})=0  \label{d}
\end{equation}
where $\partial _{A}$ denote the derivatives with respect to the $%
\alpha^{A}. $ Assuming for the moment that (\ref{d}) could be solved for the 
$\alpha ^{A}=\alpha ^{A}(x^{a}),$ then when they are substituted into (\ref
{c}) one obtains the function 
\begin{equation}
S(x^{a},x_{0}^{a})=Z(x^{a},\alpha ^{A}(x^{a}))-Z(x_{0}^{a},\alpha
^{A}(x^{a}))=0.  \label{e}
\end{equation}
Using (\ref{d}) it is easy to see that $\partial _{a}S=\partial _{a}S^{*}$
so that again $S(x^{a},x_{0}^{a})$ is a null surface thru the point $%
x_{0}^{a};$ its gradient at $x_{0}^{a}$, namely $\partial _{a}S=$ $%
Z(x_{0}^{a},\alpha ^{A}),$spans the light-cone at $x_{0}^{a}$ $[$at $%
x_{0}^{a},$ Eq.(\ref{d}) can not be solved for the $\alpha ^{A}=\alpha
^{A}(x^{a});$ all values of $\alpha ^{A}$ are allowed.] We thus see that Eq.(%
\ref{e}) represents the light-cone ${\frak C}_{x_{0}}.$ The assumption that
Eq.(\ref{d}) could be solved for $\alpha ^{A}=\alpha ^{A}(x^{a})$ depended
on the non-vanishing of the determinant $J$ of the matrix $J_{ij}\equiv
\partial _{i}\partial _{j}S^{*}(x^{a},x_{0}^{a},\alpha ^{i}).$ $J$ does
vanish at the singularities of the ``surface'' $S(x^{a},x_{0}^{a}),$ e.g.,
at the apex $x^{a}=x_{0}^{a}.$ In general, however even when $J=0$, Eqs.(\ref
{d}) and (\ref{c}) can be solved for other variables, namely {\it some }set
of{\it \ }three (say $x^{\alpha };$which might be different in different
regions) of the four $x^{a},$ in terms of the fourth one (say $x^{*})$ and
the $\alpha ^{A},$ i.e., 
\begin{equation}
x^{\alpha }=x^{\alpha }(x_{0}^{a},x^{*},\alpha ^{A}).  \label{f}
\end{equation}
Note the important point that if the coordinates $x^{a}$ are such that three
of them are space-like and one of them is a time coordinate, $x^{0},$ then
Eq.(\ref{f}) has a stronger version, namely

\begin{eqnarray}
x^{j} &=&x^{j}(x_{0}^{a},x^{*},\alpha ^{A}),  \label{f*} \\
x^{0} &=&x^{0}(x_{0}^{a},x^{*},\alpha ^{A})  \label{f**}
\end{eqnarray}
where the two $x^{j}$ and the $x^{*}$ are the three space-like coordinates.
That one can solve for the $x^{0}=x^{0}(x_{0}^{a},x^{*},\alpha ^{A})$
follows from the fact that Eq.(\ref{e}) can always be solved, from the
implicit function theorem, for $x^{0}$ since $S^{*}$ satisfies the eikonal
equation and hence $\partial S^{*}/\partial x^{0}\neq 0$

Eqs.(\ref{f*}) and (\ref{f**}) are a parametric representation of ${\frak C}%
_{x_{0}}$ via the null geodesics that rule it. For the different given
values of the $\alpha ^{A},$ they are the null geodesics thru $x_{0}^{a}.$

We thus have the result that the ${\frak C}_{x_{0}}$can be given either via
the surface (\ref{e}) or by its geodesics (\ref{f*}) and (\ref{f**}). We
will return to Eq.(\ref{f**}) later.

\begin{figure}[tbp]
\centerline{\ \hbox{\psfig{figure=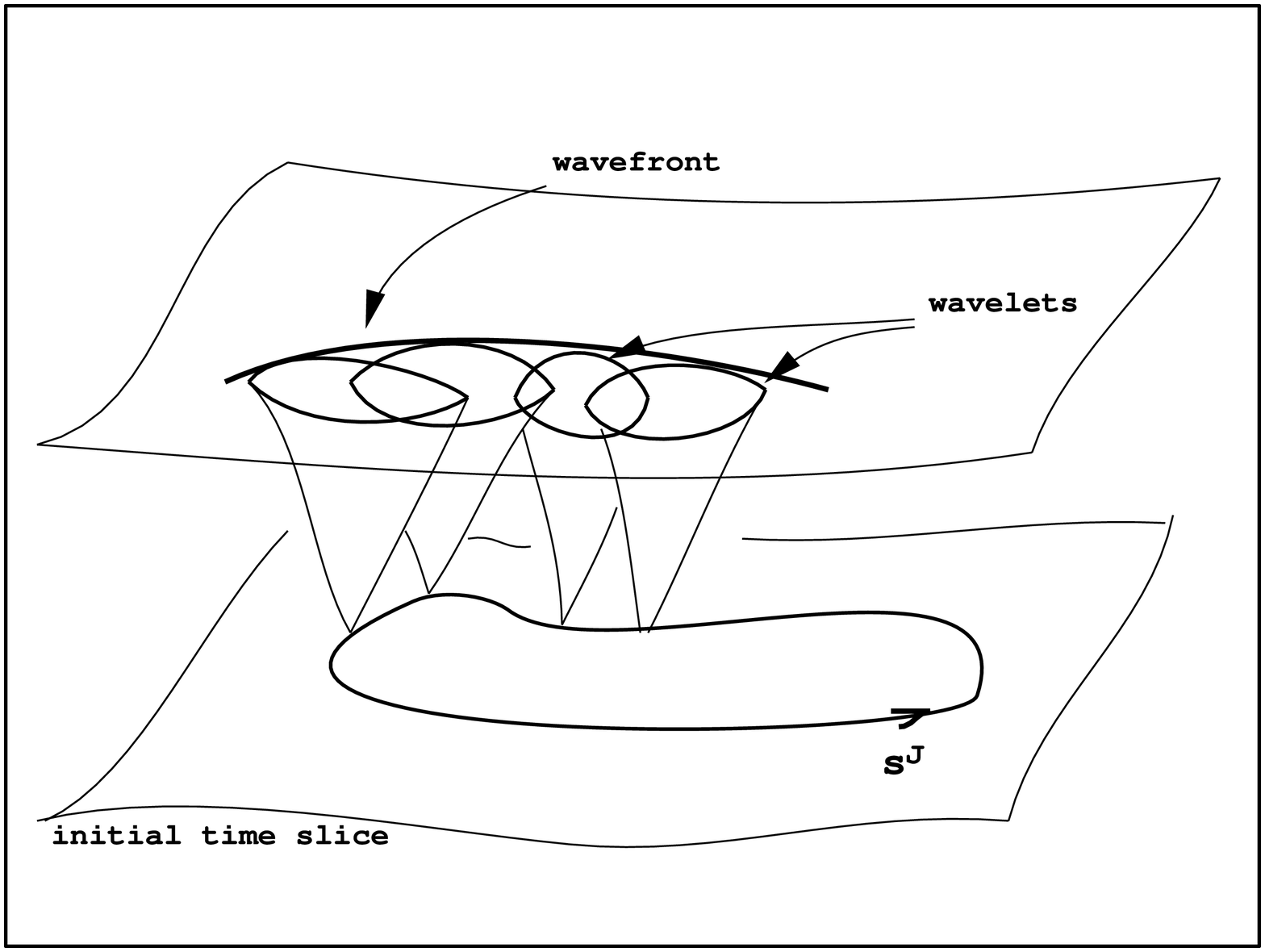,height=3in,angle=0}
	} }
\caption{The construction of the wavefront as the envelope of the individual
wavelets. }
\label{fig:huygens}
\end{figure}

If we now allow the $x_{0}^{a}$ to lie on a space-like two-surface ${\sl S}$
described by $x_{0}^{a}=x_{0}^{a}(s^{J})$, parametrized by the two
parameters $s^{J}$ , then the previous construction of light-cones yields
the family of light-cones of all the points of ${\sl S}$ via $%
x_{0}^{a}\Rightarrow x_{0}^{a}(s^{J}).$ The intersection of the set of all
the light-cones with a constant-time slice $x^{0}=$ {\it constant, }is a
family of individual (small, two-dimensional) wavefronts emanating from each
point on the surface; they are denoted as ``Huygen's wavelets''. By Huygen's
principle, the envelope of all the wavelets, at $x^{0}=$ {\it constant, }is
the two dimensional wavefront from the source ${\sl S}{\frak .}$ (see Fig.~%
\ref{fig:huygens}) The evolution, as $x^{0}$ changes, of these wavefronts
yields a new characteristic surface (big wave-front). It is equivalent to
the envelope of the family of light-cones of all the points of ${\sl S}$;
the envelope corresponding to the stationary variation of the family of
light-cones with respect to variations in the $s^{J}.$

More precisely, the envelope is the three-surface defined, first by Eqs.(\ref
{c}) and (\ref{d}), [the conditions for the light-cones from $%
x_{0}^{a}(s^{J})],$ i.e.,

\begin{equation}
S^{*}(x^{a},x_{0}^{a}(s^{J}),\alpha ^{A})=Z(x^{a},\alpha
^{A})-Z(x_{0}^{a}(s^{J}),\alpha ^{A})=0,  \label{g}
\end{equation}
\noindent

\begin{equation}
\partial _{A}S^{*}(x^{a},x_{0}^{a}(s^{J}),\alpha ^{A})=0  \label{h}
\end{equation}
augmented by the $s^{J}$ variations, i.e., by\qquad 
\begin{equation}
\partial _{J}S^{*}(x^{a},x_{0}^{a}(s^{J}),\alpha ^{A})=0.  \label{i}
\end{equation}
These are five conditions on the eight variables ($x^{a},s^{J},\alpha ^{A})$
thus forming a three-surface in the eight dimensional space; this when
projected down to the space-time results in the aforementioned envelope. It
is easily seen from Eqs.(\ref{h}) and (\ref{i}) that this surface, which we
will denote by$\qquad \qquad \qquad \qquad $%
\begin{equation}
N(x^{a})=0  \label{j}
\end{equation}
is a characteristic surface and hence satisfies the eikonal equation, (\ref
{a}). Though almost everywhere it can be given in the form of the vanishing
of a function of $x^{a},$ i.e., by Eq.(\ref{j}), there will be lower
dimensional regions where it must be given parametrically. See e.g., Eq.(\ref
{f}) or Eqs.(\ref{f*}) and (\ref{f**}).

\qquad Before looking at the time of arrival function, we first look at Eq.(%
\ref{i}) more closely. Substituting Eq.(\ref{g}) into Eq.(\ref{i}) and
taking the required derivatives we have

\begin{equation}
\partial _{a}Z(x_{0}^{a}(s^{J}),\alpha ^{A})\frac{\partial x_{0}^{a}}{%
\partial s^{J}}=0  \label{k}
\end{equation}
which is the statement that for the null ray leaving ${\sl S}$ at the point $%
x_{0}^{a}(s^{J})$, $\partial _{a}Z(x_{0}^{a}(s^{J}), \alpha ^{A})$ must be
normal to the tangent vectors $\frac{\partial x_{0}^{a}}{\partial s^{J}}$at $%
{\sl S}$ and thus normal to ${\sl S}$. Eq.(\ref{k}), hence, chooses among
all the rays forming the light-cone at $x_{0}^{a}(s^{J})$, i.e., the rays
parametrized by $\alpha ^{A},$ just the appropriate $\alpha ^{A}$ so that
the ray is the (unique) normal to ${\sl S}$. We thus have the result that ( 
\ref{k}) can be solved by

\begin{equation}
\alpha ^{A}=\alpha ^{A}(s^{J}).  \label{l}
\end{equation}
Using Eq.(\ref{l}), we have that Eqs.(\ref{g}) and (\ref{h}) become

\begin{equation}
S^{*}(x^{a},x_{0}^{a}(s^{J}),\alpha ^{A}(s^{J}))=Z(x^{a},\alpha
^{A}(s^{J}))-Z(x_{0}^{a}(s^{J}),\alpha ^{A}(s^{J}))=0,  \label{g*}
\end{equation}
\noindent

\begin{equation}
\partial _{A}S^{*}(x^{a},x_{0}^{a}(s^{J}),\alpha ^{A})|_{\alpha ^{A}=\alpha
^{A}(s^{J})}=0  \label{h*}
\end{equation}
Using the same argument that led to Eq.(\ref{f**}), namely the implicit
function theorem and $\partial S^{*}/\partial x^{0}\neq 0,$ we see that Eq.(%
\ref{g*}) is equivalent to

\begin{equation}
x^{0}=T(x^{\alpha },x_{0}^{a}(s^{J}),\alpha ^{A}(s^{J}))  \label{m}
\end{equation}

If we take the three $x^{\alpha }=X^{\alpha }$ as the ``constant spatial
position'' of the world-line of Sec. II, we have the time of arrival
function. Since, interpreting Eq.(\ref{g*}) as defining Eq.(\ref{m})
implicitly, we have that

\begin{equation}
\frac{\partial S^{*}}{\partial x^{0}}\partial _{J}T+\partial _{J}S^{*}=0,
\label{n}
\end{equation}
which, since $\partial S^{*}/\partial x^{0}\neq 0,$ implies that

\begin{equation}
\partial _{J}S^{*}=0\Rightarrow \partial _{J}T=0.
\end{equation}
Thus the extremization of $S^{*}$ implies the extremization of $T(x^{\alpha
},x_{0}^{a}(s^{J}),\alpha ^{A}(s^{J}))$ as was to be proved. This proof is
not affected by the difficulties in Sec. II of the possible multivaluedness
of the earlier $T.$

In the terminology of Arnold, these results follow from his theory of
Legendre submanifolds and maps, where

\begin{equation}
Z(x^{a},\alpha ^{A})-Z(x_{0}^{a}(s^{J}),\alpha ^{A})=0
\end{equation}
from Eq.(\ref{g}), defines a generating family $F$($x^{a},\alpha ^{A},s^{J})$
and Eqs.(\ref{h}) and (\ref{i}) define the Legendre map.


\section{Discussion \label{sec:4}}


We have given two derivations of a variational principle for the time of
arrival of null geodesics at an observer. Superficially, it appears as if it
were a version of Fermat's principle; in actuality it is quite different.
Fermat's principle leads to local evolutionary laws for the rays while here
we have from the start assumed that the rays are given by null geodesics.
Our variational principle gives the initial direction of the ray. Often
Fermat's principle is invoked to derive the equations of gravitational
lensing~\cite{faraoni,perlick}. A paper is now in preparation, using the
techniques discussed here, in which a universal lensing equation valid in
all situations is obtained.

\acknowledgements

This work received the support of the NSF grant No. PHY-9722049.


\end{document}